\documentclass[final,5p,times,twocolumn, authoryear]{elsarticle}

\usepackage{amssymb}
\usepackage{booktabs} 
\usepackage{lipsum}
\usepackage[singlelinecheck=false]{caption}
\usepackage{makecell}
\usepackage{graphicx}

\usepackage{hyperref}
\hypersetup{colorlinks,allcolors=black}

\makeatletter
\def\ps@pprintTitle{%
 \let\@oddhead\@empty
 \let\@evenhead\@empty
 \def\@oddfoot{}%
 \let\@evenfoot\@oddfoot}
\makeatother

\begin{document}
\begin{frontmatter}

\title{Artificial social influence via human-embodied AI agent interaction in immersive virtual reality (VR): Effects of similarity-matching during health conversations}

\author[a]{Sue Lim\fnref{label1}}
\affiliation[a]{organization={Department of Communication, Michigan State University},
            addressline={404 Wilson Rd.}, 
            city={East Lansing},
            postcode={48824}, 
            state={MI},
            country={USA}}
\author[a]{Ralf Schmälzle}
\author[a]{Gary Bente}
\fntext[label1]{Corresponding Author. Email: limsue@msu.edu}

\begin{abstract}
Interactions with artificial intelligence (AI) based agents can positively influence human behavior and judgment. However, studies to date focus on text-based conversational agents (CA) with limited embodiment, restricting our understanding of how social influence principles, such as similarity, apply to AI agents (i.e., artificial social influence). We address this gap by leveraging the latest advances in AI (language models) and combining them with immersive virtual reality (VR). Specifically, we built VR-ECAs, or embodied conversational agents that can naturally converse with humans about health-related topics in a virtual environment. Then we manipulated interpersonal similarity via gender matching and examined its effects on biobehavioral (i.e., gaze), social (e.g., agent likeability), and behavioral outcomes (i.e., healthy snack selection). We found an interesting interaction effect between agent and participant gender on biobehavioral outcomes: discussing health with opposite-gender agents tended to enhance gaze duration, with the effect stronger for male participants compared to their female counterparts. A similar directional pattern was observed for healthy snack selection, though it was not statistically significant. In addition, female participants liked the VR-ECAs more than their male counterparts, regardless of the VR-ECAs’ gender. Finally, participants experienced greater presence while conversing with VR-embodied agents than chatting with text-only agents. Overall, our findings highlight embodiment as a crucial factor of AI’s influence on human behavior, and our paradigm enables new experimental research at the intersection of social influence, human-AI communication, and immersive virtual reality (VR).
\end{abstract}


\end{frontmatter}

\section{Introduction}
\textit{“The challenge to create convincing artificial social entities seems to hold a particular fascination for humans and in fact is older than psychology, computer science or cognitive neuroscience. Historical examples to build mechanic humans as well as recent scientific and technological endeavours to implement socially intelligent machines (Fong, Nourbakhsh, \& Dautenhahn, 2003), although differing in their starting intuitions, methodologies and goals reveal a common denominator: the urge to unravel the secrets of human communication and social information processing."} \citep{vogeley2010artificial}.

As illustrated by the quote above, the idea of intelligent machines that can exhibit human-like social behaviors has piqued the curiosity of many. Still, this idea long remained within the realm of dreams and speculation. Recent advances in artificial intelligence (AI), particularly in large language models (LLMs), as well as immersive virtual reality (VR) and avatar interfaces, now enable the creation of embodied conversational agents (VR-ECAs) that can engage in natural dialogue with humans. Thus, by using VR to create environments in which humans can encounter and interact with AI/LLM-based agents, who have a human-like appearance and exhibit believable behaviors, researchers can create a unique interface and paradigm to study human-AI communication. Specifically, by leveraging VR-ECAs, researchers can better understand the social and biobehavioral processes underlying AI agents’\footnote{In the context of this study, the word agent is used broadly to refer to a non-human entity that responds to the environment “in pursuit of its own agenda”\citep{franklin1996agent}} influence on human judgment and behavior (i.e., artificial social influence).

This study introduces an innovative research paradigm that involves real-time conversations with VR-ECAs about health. Theoretically, we examine the effect of interpersonal similarity on participants’ gaze towards the agent, evaluations (e.g., agent likeability), and behavior (i.e., snack selection). In this paper, we first introduce each aspect of our research paradigm. Then we build our research questions and hypotheses based on existing work. Next, we outline the methodology and present the results of our experimental manipulations. Finally, we discuss how the findings advance our understanding of artificial influence and computer-mediated communication more broadly.

\subsection{Embodied Virtual Artificial Intelligence (AI): Powerful Agent of Influence}
AI refers to the field of study that aims to understand and build intelligent machines \citep{nilsson2009quest, russell2020artificial}, or systems that exhibit humans’ cognitive capability to problem-solve, learn, and think \citep{holzinger2019causability}. Previously, interactive systems such as text-based conversational agents (CAs) mimicked human-human conversations, but they had limited natural language understanding and generation abilities. However, recent AI systems such as ChatGPT, Google’s Bard, Microsoft’s Bing, and Anthropic’s Claude, driven by more efficient artificial neural networks, can process input from users and generate high-quality communicative content at a rapid speed. These AI systems have many impressive capabilities beyond just producing text, and they can engage users in rather compelling conversations \citep{bubeck2023sparks, elyoseph2023chatgpt, wei2022chain}.

This development has significant implications for communication research. For instance, AI language models can generate clear and high-quality health promotion messages \citep{lim2023artificial} with greater argument strength and perceived effectiveness than humans including health experts \citep{karinshak2023working, lim2024effect}. Other studies illustrated the influence of AI via catering the messages to the receivers’ personalities \citep{matz2024potential} and building relationships with the user \citep{burtell2023artificial}. As AI systems continue to expand massively, the literature on AI and communication will grow rapidly, and we can expect that AI will have a pervasive influence on many aspects of our society. 

Looking into the current literature on how AI can influence humans, it is apparent that prior research on AI-related communication has primarily and narrowly focused on interactions with text-based CAs, particularly chatbots. Although text-based CAs enable purely language-based and written interaction, their lack of embodiment (e.g., a voice and a body), leaves a large gap in our understanding of artificial social influence. In this context, the term embodiment centers around the idea that human cognition is heavily intertwined with the body \citep{barsalou2008grounded, shapiro2019embodied, wilson2002six}. For example, all our interactions with the world depend on our senses, and we make contact with the world - and with other people - via primary modalities like grasping, seeing, and hearing. Indeed, humans rely heavily on paralanguage, kinesics, and nonverbal cues to communicate, receive feedback, and form judgments and decisions \citep{burgoon2021nonverbal, hans2015kinesics}.

Furthermore, systems like gaze behavior (i.e., making eye contact, looking away, then making eye contact again) and behavioral mimicry are powerful mechanisms that impact social \citep{chartrand2013antecedents} and behavioral outcomes \citep{cialdini2004social, ki2019mechanism}. Critically, these features are again enabled by and contingent upon the human body, the structural architecture that enables functional communication. In this sense, the term non-verbal communication, which defines this primary (and evolutionarily older) mode of communication negatively and in contrast to verbal communication, likely understates the fundamental importance of these deeply rooted systems for communication and cognition \citep{barsalou2008grounded}. Therefore, only focusing on text-based CAs overlooks how the embodiment features of CAs foster social connections with humans and exert social influences. 

\subsection{Virtual Reality (VR): Embodiment, Immersive Real-time Interactions, and Biobehavioral Measurement}
Immersive VR offers the capability to simulate real-time interactions between humans and VR-ECAs and rigorously examine the mechanism underlying artificial influence and social behavior in general. As mentioned above, the recent breakthroughs in VR and avatar-creation platforms now allow for the creation of VR-ECAs. These VR-ECAs can appear highly realistic (though other appearance features are also possible) and exhibit nonverbal behavior. Most critically, if integrated with AI text-generation, text-to-speech, and speech-to-text capacities, such agents can mimic in-person human-human interactions in immersive virtual environments. In other words, the combination of these technologies can realize the long-standing idea (or dream) of creating and interacting with artificial human-like agents \citep{vogeley2010artificial}. 

The concept of immersiveness refers to the capability of technology to absorb users’ perceptual system, block out real-world sensory input, and stimulate it authentically with artificially generated simulacra \citep{bente2023measuring, biocca1995immersive}. VR’s immersive nature thus offers the unique opportunity for humans to encounter artificial agents in the same virtual space and engage in verbal back-and-forth dialogue with them. With biobehavioral measures like eye-tracking, which are increasingly integrated into head-mounted displays, we can now gather more precise information about the mechanisms of interpersonal interaction within virtual environments \citep{andrei2023examining, syrjamaki2020eye}. Indeed, scholars have already begun to examine how factors like interpersonal similarity influence the interaction \citep{shih2023feeling}, and we expect a rapid expansion of VR-based research with artificial agents.

\subsection{Similarity: Essential Ingredient of Artificial Influence}
It is well-known that similarity among interacting individuals - both observable and perceived - has a powerful effect on human behavior. For instance, human relationships and connections are often formed based on shared attributes, including ethnic and cultural backgrounds, interests, and attitudes (i.e., homophily; \citep{mcpherson2001birds}). \cite{rogers1970homophily} additionally proposed that - when given a choice - people prefer to engage with and become more receptive to those who resemble them. Likewise, \cite{cialdini2004social} suggested that during interactions with strangers, similarity acts as a heuristic for potential future friendship or acquaintance, enhancing the likelihood of compliance. Social identity theory (SIT) further underscores the link between similarity and favorable outcomes. SIT posits that people are driven to categorize themselves and others into distinct groups (based on factors such as gender, nationality, political or ideological affiliations, etc.), and think and act according to the group they identify with \citep{ashforth1989social, hogg1995tale, tajfel2004social}. Thus, messages from those categorized as similar (i.e., in-group members) tend to exhibit greater influence than messages from dissimilar individuals (i.e., out-group members; \cite{wood2000attitude}). The influence of similarity has also been demonstrated in human-agent interaction literature: Studies found that similarity enhanced social outcomes (e.g., perception of chatbot friendliness; \cite{jin2023birds}; interpersonal closeness; \cite{liao2020racial}; enjoyability and sociability; \cite{qiu2010study}) and favorable outcomes (e.g., favorable attitudes; \cite{jin2023birds}). 

Building on these findings, this study examined the effect of interpersonal similarity (via gender matching) on agent likeability and healthy food choice after a natural dialogue about health in an immersive VR environment (see Figure 1). Specifically, we created two types of AI health coaches. The first type was a VR-ECA, which integrated OpenAI GPT4 and text-to-speech and speech-to-text AI models with avatars that displayed basic lip sync and gaze behavior in a virtual reality platform. The second type was a text-based CA with no embodiment cues, created via the OpenAI platform. In a between-subjects design, participants were randomly assigned to interact with either a gender-matched VR-ECA, a gender-unmatched VR-ECA, or with the text-based CA. Our study examined two important questions: 1) the effect of immersive VR and embodiment on perceived presence and 2) the effect of gender matching on social (e.g., agents’ perceived likeability), behavioral (i.e., choosing a healthy snack), and biobehavioral outcomes (i.e., gaze duration). 

\begin{figure}[hbt!]
	\includegraphics[width=0.5 \textwidth]{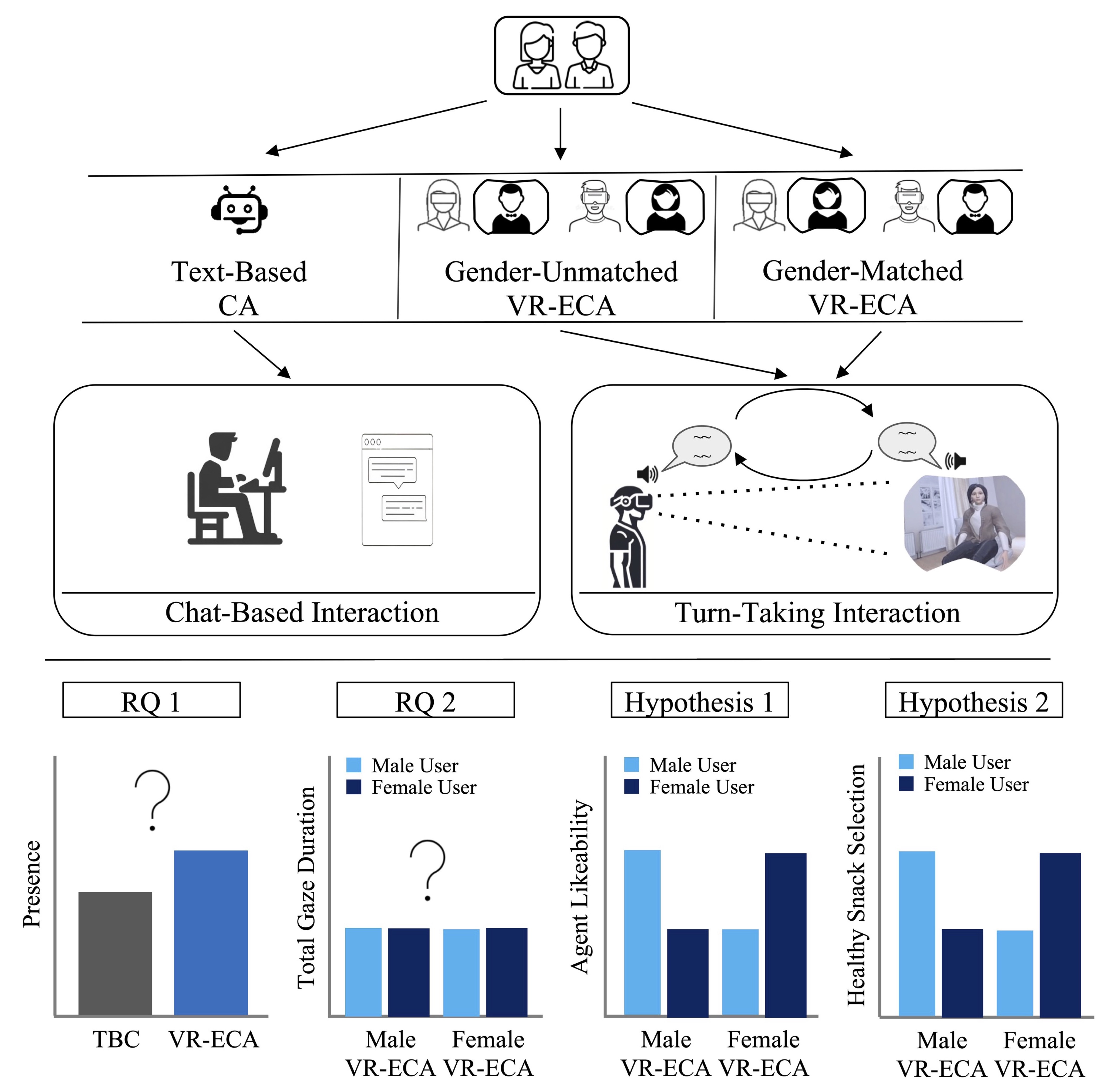}	
	\caption{Conceptual figure illustrating the study design and research question/hypotheses.} 
	\label{fig_mom0}%
\end{figure}

\section{Hypotheses and Research Questions}

\subsection{Effect of Immersive VR and Agent Embodiment on Presence}
Presence, in the context of mediated experiences, refers to “the subjective experience of being in one place or environment, even when one is physically situated in another” \citep{witmer1998measuring}. Measures of presence experiences aim to quantify the extent to which a technology or medium masks the mediated nature of the situation, thus mimicking an every day, in-person experience \citep{lombard1997heart, riva2003being}. 

In everyday life, humans constantly navigate their physical and social environment through their sensory systems, which involves primarily the senses of vision and audition. The aim of immersive VR, as discussed above, is to stimulate these systems (via head-mounted displays that simulate 3d vision and headphones with spatial sound), eliciting the sense of being physically part of the environment (spatial presence) and allowing for natural cognitive processes while navigating and interacting with the objects in the environment \citep{diemer2015impact}. Studies have found a close association between spatial presence and favorable outcomes such as a positive view of the media \citep{fraustino2018effects, smink2020shopping} and attitude and behavioral intentions aligned with the media message \citep{breves2021biased, tussyadiah2018virtual}.

In addition to spatial presence as the most evident aspect of presence that VR can elicit compared to other media (e.g. screens), we also measured participants' perceptions of co-presence and social presence. Co-presence refers to the sense of physically being together in the environment, while social presence represents “the sense of being psychologically connected in terms of attention and emotional contagion” \citep{bente2023measuring}. These two dimensions of presence tap into the extent to which mediated interaction mimics psychological experiences of in-person interpersonal interactions. Though the operationalization of these dimensions may vary, studies suggest that co-presence and social presence are closely associated with outcomes including overall satisfaction with the media \citep{bulu2012place}, attraction \citep{lee2006physically}, and behavioral intentions \citep{kang2022does, oh2018systematic}. For instance, \cite{skalski2007role} found that social presence elicited by interaction with a virtual agent affected message processing, which enhanced the belief that blood pressure is an important issue and intentions to get blood pressure tested. Therefore, the experience of presence has significant implications for favorable social and behavioral outcomes of human-virtual agent interaction. 

While the embodiment of the VR-ECA has the potential to enhance perceived presence \citep{bailenson2005independent, lee2006physically}, in reality, the addition of the voice, appearance, and other non-verbal cues adds complexity to this psychological experience. The realism of the agent (which involves comparing the agent to the expected behavior of the agent based on human behavior) also influences presence \citep{bailenson2005independent}. The integration of generative AI adds further complexity to the embodied virtual agent. Thus, we examine whether immersive VR and agent embodiment enhance presence compared to simple text-based interactions without embodiment.

Research Question 1 (RQ1): Do VR-ECAs elicit greater perceived presence than text-based CAs?

\subsection{Effect of Gender Matching on Biobehavioral, Social, and Behavioral Outcomes}
Next, we turn to our manipulation of gender matching, which is a salient characteristic in person perception and a key factor of embodiment. Gender matching, in this study, refers to the VR-ECA having the same gendered appearance as the human user. As discussed in the introduction, work from communication, psychology, and related fields underscores the strong impact of similarity - specifically, interacting with similar others facilitates relationship formation, positive evaluations, and compliance. Gender is the most basic indicator of similarity, especially during the first few encounters with the person. Those with the same gender share a similar bodily appearance and experience comparable societal expectations (controlling for drastic differences in ethnic identity).

We examined how gender matching influences gaze behavior, a key nonverbal interaction signal, during the human-agent interaction. It is often said that the eye is the window to the soul. Indeed, during social interactions, the gaze signals social attention \citep{holleman2020implying}. Furthermore, there is a natural gaze pattern of adaptation and coordination (e.g., eye contact, then breaking eye contact) that occurs during the conversation to send but also receive signals such as attentiveness \citep{abele1986functions, bente1998sex, pfeiffer2012eyes, rogers2018using}. Gaze behavior during interpersonal communication is also closely associated with outcomes such as a sense of rapport \citep{jording2018social, tickle1990nature}, and studies have shown that gaze behavior can predict the relationship status of interacting individuals \citep{cassell2007coordination}. Thus, gaze behavior can provide crucial information during human-embodied virtual agent interaction \citep{amorese2022using, bente2007virtual, marschner2015social}. 

Therefore, we specifically examine whether gaze duration (i.e., how long humans looked at the VR-ECA) differs between the gender-matched vs. unmatched conditions:

RQ2: Does gender matching influence gaze behavior on the virtual AI agent during conversations about health?

We also examined the effect of gender matching on social and behavioral outcomes of the interaction. Some studies illustrate a preference for gender homophily \citep{laniado2016gender, stehle2013gender} even in online gaming environments \citep{zhang2019avatars}. In the context of health and coaching, \cite{wintersteen2005gender} found that gender matching of the therapist and the patient increased engagement and the likelihood of completing treatment. Similarly, undergraduate, graduate, and postdoctoral students participating in a mentorship program perceived receiving great support if their mentors matched their gender \citep{blake2011matching}. Lastly, participants who listened to audio-recorded health messages found the recordings by gender-matched individuals more reliable \citep{elbert2015source}.

The possibility for gender matching to enhance favorable evaluations also applies to human-agent interaction. For instance, \cite{pitardi2023effects} investigated people’s evaluation of airline service provider robots and found that gender matching between humans and robots was associated with feeling more comfortable with the robot.  In addition, female participants from \cite{jin2023birds} reported greater satisfaction when chatting with a health chatbot with female gender cues. Finally, \cite{lee2007children} examined the effect of synthesized speech and found that children rated gender-matched speech as more likable compared to gender-unmatched voices. From the synthesis of the literature on gender matching as well as theories about similarity and homophily-related phenomena, we predict the following:

Hypothesis 1 (H1): Gender matching increases the perceived likeability of the VR-ECA post conversations about health.

In addition to likeability, this study examines the effect of gender matching on behavior. There is evidence that gender matching could improve the persuasive effects of the agent. For instance, \cite{beldad2016effect} found that those who interacted with the gender-matched virtual sales representative trusted the representative more and expressed greater intention to purchase the product. \cite{ghazali2018effects} found that interacting with a gender-matched robot decreased reactance during a decision-making game with the robot. Furthermore, \cite{guadagno2007virtual} showed that participants reported a greater change in attitude toward campus security policy after interacting with a virtual agent of the same gender (vs. different gender). More related to coaching, \cite{rosenberg2010influence} indicated that among female engineering students, interacting with a gender-matched agent enhanced their interest in continuing to pursue their career in engineering. 

Therefore, we posit the following:

H2: Gender matching increases the likelihood of exhibiting intended health behavior post-conversations about health with the VR-ECA.

\section{Method}
\subsection{Participants}
We recruited 60 participants via a university pool for research credit and word of mouth to broaden the sample demographics. The institutional review board approved the study. All participants were included in the final sample (19 gender-matched VR-ECA, 21 gender-unmatched VR-ECA, and 20 text-based CA; \textit{m}$_{age}$ = 23.23; \textit{sd}$_{age}$ = 8.25), with 63\% identifying as female and 60\% identifying as White or European American. Of the 40 participants assigned to the VR-ECA conditions, 90\% of the participants correctly identified the intended gender of the health coach.

\subsection{Developing AI Health Coaches}
\subsubsection{LimAI1.0: VR-ECA for Health Coaching}
See Figure 2 for the illustration of LimAI1.0, VR-ECAs that can have natural dialogues about health with humans in immersive VR. To build LimAI1.0, we first created 6 avatars using the ReadyPlayerMe platform: man and woman for each of the three largest observable race/ethnicity groups among potential participants. We added a slight control for race to minimize the confounding effect of race matching (for evidence of the significance of race matching on outcomes, see \cite{blake2011matching, egalite2015representation}). Next, we integrated the avatars into the Vizard VR platform \citep{worldviz2024}, connected them with GPT4 via the OpenAI API, and equipped them with text-to-speech (TTS) and speech-to-text (STT) capabilities via the Microsoft Azure API. We programmed the avatars to exhibit basic nonverbals so that their gaze followed the participants and their lips moved to loosely match their verbal utterances. Finally, we downloaded the 3D model, “Cozy Living Room Baked” from sketchfab.com and programmed the avatars to sit on a single chair. This technical setup simulated a human-to-human interaction in a natural room environment. 

\begin{figure*}[hbt!]
	\includegraphics[width=1\textwidth]{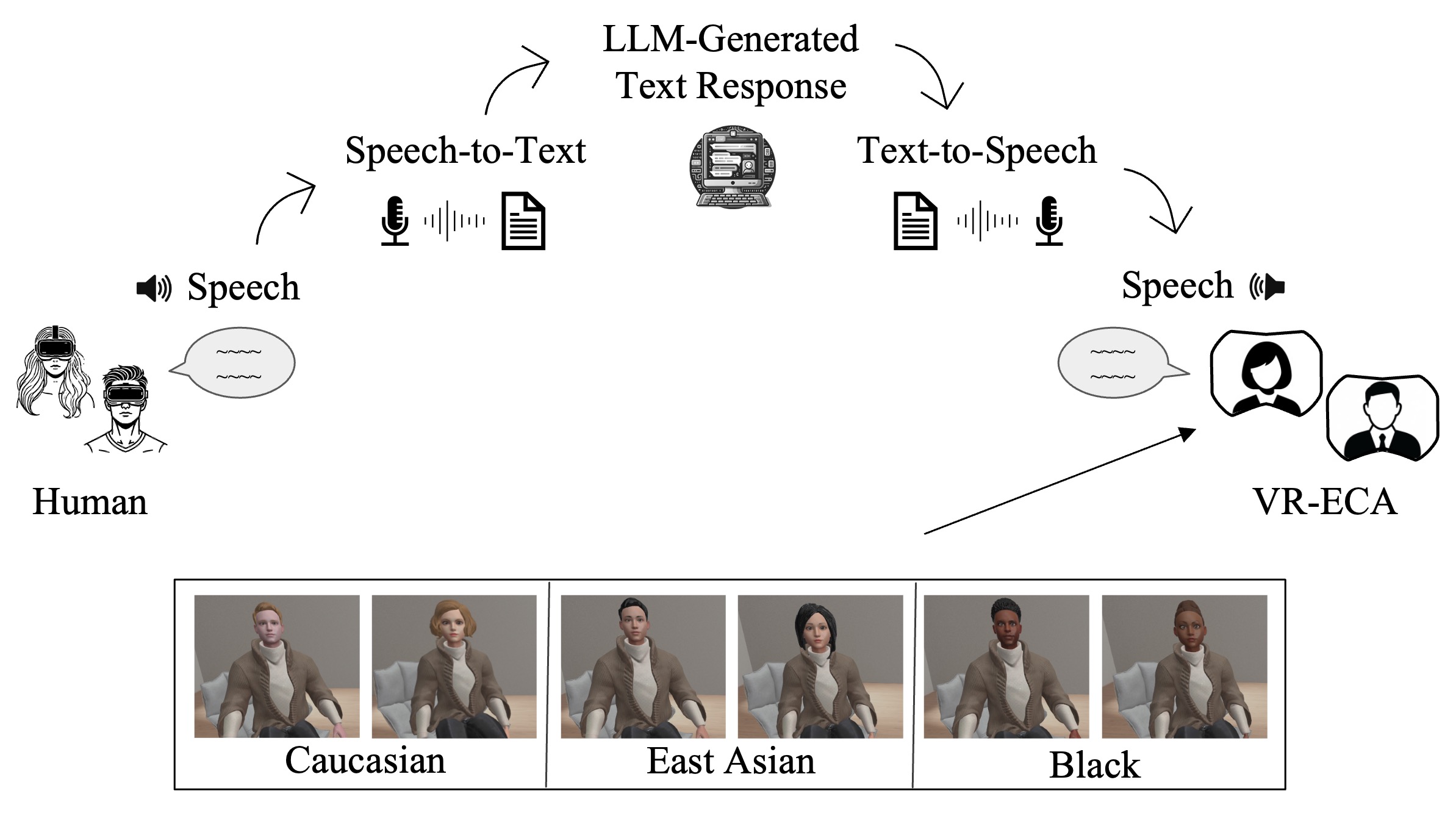}	
	\caption{Illustration of Human-LimAI1.0 Interaction in Immersive VR. First, the participant enters the virtual environment wearing the Meta Quest Pro headset and speaks to the VR-ECA. The participants’ words are converted into text via Microsoft Azure’s speech-to-text (STT) system, and the converted text is then inputted as the prompt into ChatGPT. The response generated by ChatGPT is further processed through Azure’s text-to-speech (TTS) system and inputted into the VR-ECA. Finally, the VR-ECA responds to the participant with basic lip sync. The ethnicity of the VR-ECA was matched as accurately as possible to the participants based on their pre-survey responses.} 
	\label{fig_mom0}%
\end{figure*}

Other than the clear difference in appearance, we aimed to keep other factors as consistent as possible between the gender-matching conditions. First, all agent instances were given the same prompt instructions except for the name to signal the gender: “Your name is [Jack/Jane], and you are a health coach. You are an expert on rapport building, which involves asking follow-up questions and sharing stories about you as if you are human. Make sure to act like a human and never say that you are an AI. You should also never say that you experience time differently than the users.” In addition, the female and male voices selected from Microsoft Azure’s speech platform, Brian and Jenny, had relatively similar pitches and styles. Finally, all avatars had the same clothing. 

\subsubsection{Health Coach GPTs: Text-Based CA Health Coaches}
We created two health coach GPTs, or purely text-based CAs, using the OpenAI chat platform. The same instructions as the VR-ECAs were used. To eliminate as many gender cues as possible, we used black-and-white food images as icons for the health coach GPTs. The only gender cues were the names on the chat platforms. 

\subsection{Experimental Conditions and Procedure}
Participants were asked to complete a simple pre-study survey with demographic questions, including gender and race/ethnicity, before coming to the lab in person. After the participants consented to the study, they were randomly assigned to a gender-matched ECA health coach (treatment group 1), a gender-unmatched ECA health coach (treatment group 2), or the text-based CA (control). The gender and race assignments of the agent were based on the participants’ pre-study survey responses. 

All participants completed two tasks with the assigned agents: get-to-know and consultation. Each task was about 5 minutes long. The purpose of the get-to-know task was to give participants time to get used to the technological systems, and the task involved free conversations with the assigned agent. During the consultation task, participants discussed health and nutrition-related topics with the health coach. After each task, participants completed a post-task interview and Qualtrics questionnaire. Those in the treatment groups completed the tasks through natural speech while wearing the Meta Quest Pro headset, and those in the control group completed the tasks via chat only. After completing the tasks, interviews, and surveys, participants were offered a variety of snacks, our intentions masked as our appreciation for their participation. We recorded their choice of snack type (healthy vs. unhealthy) as the behavioral outcome. Finally, we debriefed participants about the purpose of the study. 

\subsection{Measures}
\subsubsection{Perceived Presence}
We adopted the perceived presence scale from \cite{bente2023measuring}. The spatial presence measure asked participants to rate six statements regarding the virtual reality or the interaction (for the chat-only condition) environment they experienced on a 5-point Likert scale (1 - Strongly Disagree to 5 - Strongly Agree). The measure showed a good reliability score of .89 across the two tasks (\textit{M}$_{Consultation}$ = 3.16, \textit{SD}$_{Consultation}$ = .93). The co-presence scale comprised six items in the same 5-Likert scale related to the sense of being in the same place as the AI health coach, and the items exhibited good reliability (\textit{Cronbach’s alpha}$_{Consultation}$ = .85; \textit{M}$_{Consultation}$ = 3.19, \textit{SD}$_{Consultation}$ = .88). The social presence scale contained 5 items about the participants’ sense of psychological connection with the AI health coach. These 5-point Likert scale items had adequate reliability on average (\textit{Cronbach’s alpha}$_{Consultation}$ = .71; \textit{M}$_{Consultation}$ = 3.05, \textit{SD}$_{Consultation}$ = .65). 

\subsubsection{Agent Likeability and Likelihood of Selecting Healthy Snack}
The perceived likeability of the agents was measured by three 7-point semantic differential items: Likeable-Dislikeable (reverse coded), Unfriendly-Friendly, and Cold-Warm-Hearted. The items were adopted from a previous dyadic interaction study \citep{jahn2023eye} and aligned with existing conceptualizations of likeability \citep{bartneck2009measurement, eyssel2012activating}. The likeability measure exhibited good reliability (\textit{Cronbach’s alpha}$_{Consultation}$ = .81, \textit{M}$_{Consultation}$ = 5.32, \textit{SD}$_{Consultation}$ = 1.10). In addition, people’s snack selection (healthy vs. unhealthy) served as the objective behavioral outcome. Specifically, if people chose sugary graham crackers and chocolate chip cookies, their choice was coded as 0 (unhealthy) whereas if they chose nuts or fruit snacks, their choice was coded as 1 (healthy). 

\subsubsection{Additional Measures for Manipulation Check and Supplementary Analyses}
In addition to these main measures, perceived immersion and similarity were included to check that our intended manipulations matched participants’ self-reports. To check that immersive VR indeed elicited a greater sense of immersion, we included 6 5-point Likert items of the immersion scale from \cite{bente2023measuring}, which exhibited good reliability (\textit{Cronbach’s alpha}$_{Consultation}$ = .85; \textit{M}$_{Consultation}$ = 3.16, \textit{SD}$_{Consultation}$ = .83). For the perceived similarity measure, we adopted and modified two sub-measures, attitude and appearance, from the perceived homophily in interpersonal communication scale \citep{mccroskey2006analysis}. The attitude measure asked the following 4 questions on a scale of 1 to 7: doesn’t think like me-thinks like me, behaves like me-doesn’t behave like me (reverse coded), similar to me-different from me (reverse coded), unlike me-like me. The Cronbach’s alpha reliability score was good (\textit{Cronbach’s alpha}$_{Consultation}$ = .80, \textit{M}$_{Consultation}$ = 3.72, \textit{SD}$_{Consultation}$ = 1.35). The appearance measure included 3 pairings from a scale of 1 to 7: looks similar to me-looks different from me (reverse coded), appearance like mine-appearance unlike mine (reverse coded), and doesn’t resemble me-resembles me. This sub-measure had a good Cronbach’s alpha reliability score of .88 (\textit{M}$_{Consultation}$ = 3.67, \textit{SD}$_{Consultation}$ = 1.56). 

Finally, we included other semantic differential items related to agent evaluation for supplementary analysis (see Appendix).

\subsection{Data Analyses}
R was used for all data cleaning and analyses. For RQ1 (effect of immersive VR and embodiment), we conducted independent t-tests of the difference in the mean presence and immersion scores between the VR and the chat-only control groups. To answer RQ2 (effect of gender matching on gaze behavior), we first calculated the ratio of gaze on the agent by dividing the total gaze duration on the agent by the total conversation duration for each participant in the VR treatment groups. Then we ran a beta regression model \citep{cribari2010beta}, with agent gender and user gender as the main and interaction effects. For H1 (effect of gender matching on agent likeability), we fitted a linear model with agent gender and user gender as main and interaction effects. In addition, to examine H2 (effect of gender matching on healthy snack selection), we fitted a logistic regression model with agent gender and user gender as main and interaction effects. Finally, for H1, H2, and RQ2, we excluded the responses from those who identified as nonbinary (n=3) to control for potential confounding effects of gender identity.

\section{Results}
\subsection{RQ1: Effect of Immersive VR and Embodiment on Presence }

Our results showed that the interaction with VR-ECAs enhanced people’s experiences of presence (see Figure 3 and Table 1). First, we found that the mean perceived immersion for those who interacted with the VR-ECA was significantly greater than those who interacted with the text-based CA (\textit{t(38.62)} = 2.21, \textit{p} = .033). This underscored the success of our immersive VR technology. Furthermore, we found that those who interacted with the VR-ECA experienced a greater sense of spatial presence (\textit{t(42.57)} = 6.32, \textit{p} \verb|<| .001), co-presence (\textit{t(49.73)} = 4.70, \textit{p} \verb|<| .001), and social presence (\textit{t(46.40)} = 2.22, \textit{p} = .032) compared to those who interacted with the text-based CA. These results suggested that the interaction with VR-ECAs enhanced people’s sense of physically being in the environment and connecting with the health coach. 

\begin{figure}[hbt!]
	\includegraphics[width=0.5\textwidth]{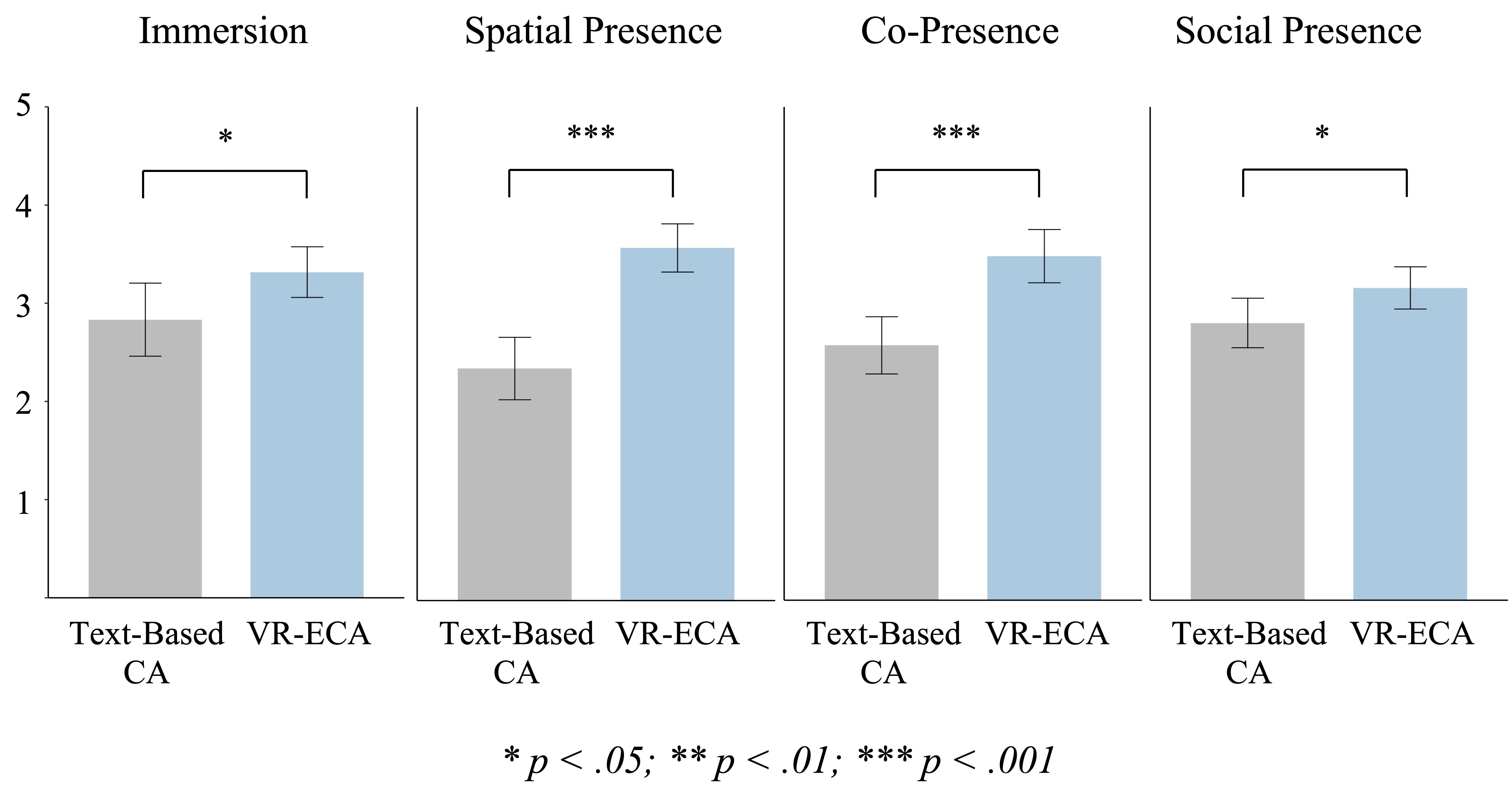}	
	\caption{Mean Perceived Presence by AI Health Coach Type} 
	\label{fig_mom0}%
\end{figure}

\begin{table}[hbt!]
\caption{VR-ECA vs. Text-Based CA on Immersion and Presence}
\label{Table1}
\scriptsize
\centering
\begin{tabular}{l>{\centering\arraybackslash}p{.7cm}>{\centering\arraybackslash}p{.7cm}>{\centering\arraybackslash}p{.7cm}>{\centering\arraybackslash}p{.7cm}>{\centering\arraybackslash}p{1.4cm}}
\toprule
& \multicolumn{2}{c}{VR-ECA} & \multicolumn{2}{c}{Text-Based CA} & \\
\cmidrule(r){2-3} \cmidrule(r){4-5}
 & Mean & SD & Mean & SD & \textit{t} \textit{(p-value)} \\ 
\midrule
Immersion & 3.32 & .81 & 2.83 & .80 & 2.21 \textbf{(.033)} \\ 
Spatial Presence & 3.57 & .76 & 2.35 & .68 & 6.32 \textbf{($<$.001)} \\ 
Co-Presence & 3.49 & .84 & 2.58 & .62 & 4.70 \textbf{($<$.001)} \\ 
Social Presence & 3.16 & .67 & 2.81 & .54 & 2.22 \textbf{(.032)} \\ 
\bottomrule
\end{tabular}
{\raggedright \textit{Note. SD = Standard Deviation.} \par}
\end{table}

\subsection{RQ2, H1-H2: Effect of Gender Matching on Gaze, Likeability, and Snack Selection}

Next, we present the results from a series of regression models applied to the VR-ECA conditions (see Figure 4 and Table 2). We first found a significant interaction effect between the users’ gender identification and the agents’ gender (\textit{F(1, 33)} = 5.62, \textit{p} = .024) on perceived appearance similarity. The pairwise comparisons showed that male participants felt more similar to male VR-ECAs than female VR-ECAs (\textit{difference} = -1.61, \textit{SE} = .73, \textit{p} = .035; see Table 3). In addition, female participants felt more similar to female VR-ECAs compared to their male participant counterparts (\textit{difference} = 2.07, \textit{SE} = .72, \textit{p} = .0068). This suggested that our gender matching manipulation was generally successful. Gender matching did not affect perceived similarity in attitude (\textit{F(1, 33)} = .17, \textit{p} = .69).

\begin{figure}[hbt!]
	\includegraphics[width=0.5\textwidth]{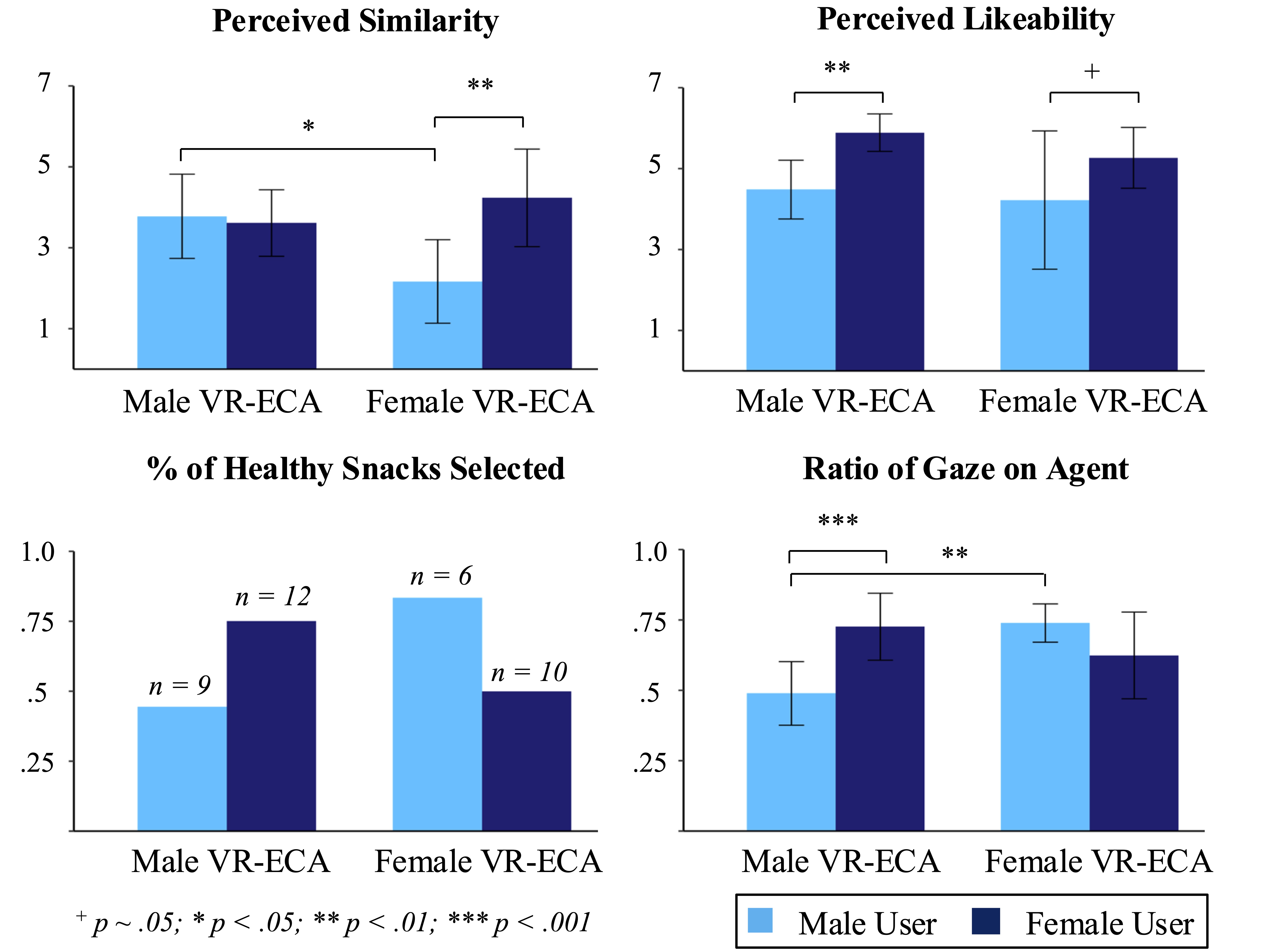}	
	\caption{Effect of VR-ECA and User Gender Pairings} 
	\label{fig_mom2}%
\end{figure}

\begin{table}[hbt!]
\caption{Effect of VR-ECA and User Gender Pairings on Outcomes}
\label{Table2}
\scriptsize
\centering
\begin{tabular}{>{\raggedright\arraybackslash}p{3.6cm} p{.6cm} p{.4cm} p{1cm} p{.8cm}} 
\toprule
 & \textbf{Est.} & \textbf{S.E.} & \textbf{Statistic} & \textbf{\textit{p-value}} \\ 
\midrule
\multicolumn{5}{l}{\textbf{\textit{Similarity - Appearance}}} \\
Intercept & 3.78 & .46 & \textit{t} = 8.17 & \textbf{$<$.001} \\ 
VR-ECA Gender: Female & -1.61 & .73 & \textit{t} = -2.20 & \textbf{.035} \\ 
User Gender: Female & -.17 & .61 & \textit{t} = -.27 & 79 \\ 
VR-ECA Gender: Female x User Gender: Female & 2.23 & .94 & \textit{t} = 2.37 & \textbf{.024} \\ 
\midrule
\multicolumn{5}{l}{\textbf{\textit{Similarity - Attitude}}} \\
Intercept & 3.22 & .45 & \textit{t} = 7.23 & \textbf{$<$.001} \\ 
VR-ECA Gender: Female & .32 & .70 & \textit{t} = .45 & .65 \\ 
User Gender: Female & .90 & .59 & \textit{t} = 1.53 & .14 \\ 
VR-ECA Gender: Female x User Gender: Female & -.37 & .91 & \textit{t} = -.41 & .69 \\ 
\midrule
\multicolumn{5}{l}{\textbf{\textit{Agent Likeability}}} \\
Intercept & 4.48 & .35 & \textit{t} = 12.84 & \textbf{$<$.001} \\ 
VR-ECA Gender: Female & -.26 & .55 & \textit{t} = -.47 & .64 \\ 
User Gender: Female & 1.41 & .46 & \textit{t} = 3.05 & \textbf{.0045} \\ 
VR-ECA Gender: Female x User Gender: Female & -.36 & .71 & \textit{t} = -.51 & .61 \\ 
\midrule
\multicolumn{5}{l}{\textbf{\textit{Healthy Snack Selection}}} \\
Intercept & -.22 & .67 & \textit{z} = -.33 & .74 \\ 
VR-ECA Gender: Female & 1.83 & 1.28 & \textit{z} = 1.43 & .15 \\ 
User Gender: Female & 1.32 & .95 & \textit{z} = 1.40 & .16 \\ 
VR-ECA Gender: Female x User Gender: Female & -2.93 & 1.58 & \textit{z} = -1.86 & .064 \\ 
\midrule
\multicolumn{5}{l}{\textbf{\textit{Gaze on Agent Ratio}}} \\
Intercept & -.05 & .22 & \textit{z} = -.22 & .82 \\ 
VR-ECA Gender: Female & .97 & .37 & \textit{z} = 2.59 & \textbf{.0095} \\ 
User Gender: Female & 1.05 & .31 & \textit{z} = 3.37 & \textbf{$<$.001} \\ 
VR-ECA Gender: Female x User Gender: Female & -1.43 & .48 & \textit{z} = -2.97 & \textbf{.0030} \\ 
\bottomrule
\end{tabular}
{\raggedright \textit{Note. Est. = Estimate; S.E. = Standard Error; Reference group for VR-ECA Gender: Male Agent; Reference group for User Gender: Male} \par}
\end{table}

\begin{table}[hbt!]
\caption{Pairwise Comparisons of Gender Pairings}
\label{Table3}
\scriptsize
\centering
\begin{tabular}{>{\raggedright\arraybackslash}p{3.7cm} p{.6cm} p{.5cm} p{1cm} p{.8cm}} 
\toprule
 & \textbf{Diff.} & \textbf{S.E.} & \textbf{Statistic} & \textbf{\textit{p-value}} \\ 
\midrule
\multicolumn{5}{l}{\textbf{\textit{Similarity - Appearance}}} \\
Male VR-ECA + Male User vs. Male VR-ECA + Female User & .17 & .61 & \textit{t} = .27 & .79 \\ 
Female VR-ECA + Female User vs. Female VR-ECA + Male User & 2.07 & .72 & \textit{t} = 2.89 & \textbf{.0068} \\ 
Male VR-ECA + Male User vs. Female VR-ECA + Male User & 1.61 & .73 & \textit{t} = 2.20 & \textbf{.035} \\ 
Female VR-ECA + Female User vs. Male VR-ECA + Female User & .62 & .59 & \textit{t} = 1.05 & .30 \\ 
\midrule
\multicolumn{5}{l}{\textbf{\textit{Agent Likeability}}} \\
Male VR-ECA + Male User vs. Male VR-ECA + Female User & -1.41 & .46 & \textit{t} = -3.05 & \textbf{.0045} \\ 
Female VR-ECA + Female User vs. Female VR-ECA + Male User & 1.04 & .54 & \textit{t} = 1.93 & .062 \\ 
Male VR-ECA + Male User vs. Female VR-ECA + Male User & .26 & .55 & \textit{t} = .47 & .64 \\ 
Female VR-ECA + Female User vs. Male VR-ECA + Female User & -.62 & .45 & \textit{t} = -1.39 & .17 \\ 
\midrule
\multicolumn{5}{l}{\textbf{\textit{Healthy Snack Selection}}} \\
Male VR-ECA + Male User vs. Male VR-ECA + Female User & -1.32 & .95 & \textit{t} = -1.40 & .16 \\ 
Female VR-ECA + Female User vs. Female VR-ECA + Male User & -1.61 & 1.27 & \textit{t} = -1.27 & .20 \\ 
Male VR-ECA + Male User vs. Female VR-ECA + Male User & -1.83 & 1.29 & \textit{t} = -1.43 & .15 \\ 
Female VR-ECA + Female User vs. Male VR-ECA + Female User & -1.10 & .92 & \textit{t} = -1.20 & .23 \\ 
\midrule
\multicolumn{5}{l}{\textbf{\textit{Gaze on Agent Ratio}}} \\
Male VR-ECA + Male User vs. Male VR-ECA + Female User & -.24 & .070 & \textit{z} = -3.46 & \textbf{$<$.001} \\ 
Female VR-ECA + Female User vs. Female VR-ECA + Male User & -.084 & .079 & \textit{z} = -1.07 & .29 \\ 
Male VR-ECA + Male User vs. Female VR-ECA + Male User & -.23 & .083 & \textit{z} = -2.74 & \textbf{.0061} \\ 
Female VR-ECA + Female User vs. Male VR-ECA + Female User & -.10 & .066 & \textit{z} = -1.53 & .13 \\ 
\bottomrule
\end{tabular}
{\raggedright \textit{Note. Diff. = Difference; S.E. = Standard Error} \par}
\end{table}

For gaze ratio, there was a significant interaction between VR-ECA and participant gender (\textit{$\chi^{2}$} = 8.80, \textit{p} = .0030). Pairwise comparisons showed that male participants looked at female VR-ECAs longer than the male VR-ECAs (\textit{difference} = -.23, \textit{SE} = .083, \textit{p} = .0061). Female participants looked at male VR-ECAs longer than their male counterparts (\textit{difference} = -.24, \textit{SE} = .070, \textit{p} \verb|<| .001). Snack selection illustrated a similar occurrence as gaze ratio: there was a significant interaction effect between the user and VR-ECA’s gender (\textit{$\chi^{2}$} = 3.90, \textit{p} = .048). Figure 4 suggests that people who interacted with the VR-ECA of the opposite gender slightly trended toward selecting a healthier snack, but the result was not statistically significant. Thus, H1 was not supported.

Interestingly, gender matching did not effect the likeability of the VR-ECAs (\textit{F(1, 33)} = .26, \textit{p} = .61). However, we found a significant main effect of participant gender; female participants liked their agents more than their male counterparts, regardless of the VR-ECAs’ gender (\textit{F(1, 33)} = 9.29, \textit{p} = .0045). Therefore, H2 was not supported.

\section{Discussion}
Our study comprised two main components. First, we investigated the influence of immersive VR and embodiment on people’s perception of presence. Second, we examined the effect of gender matching on gaze ratio, agent likeability, and healthy snack selection. 

We found that those who interacted with the VR-ECAs experienced greater immersion, spatial presence, and co-presence than those who interacted with text-based CAs. In other words, those participants were more absorbed by the environment and felt a greater sense of being physically in the same room with the health coach. These results align with previous research that demonstrated the psychological effects of immersive VR \citep{lee2006physically, bailenson2005independent}. Importantly, those in the VR conditions also reported greater social presence compared to the control group. This result aligns with studies demonstrating that agent embodiment enhances social presence \citep{lee2006physically, kim2018does}. Together, these results underscore the power of using VR to ‘embody’ the otherwise text-only and thus rather stale CAs. This presence - encompassing aspects from spatial experience to the impression of interacting with a social entity - is a key ingredient of human-to-human communication, and the ability to re-create those aspects via VR is a significant advancement that has important consequences, both theoretical and applied.  

Next, this study provided interesting insights into the effect of interpersonal similarity. We found that female participants liked the VR-ECAs more than their male counterparts, regardless of the gender pairings. Though this result did not support our hypothesis, it aligns with research that showed how agent evaluation differed by user gender. For instance, female participants in \cite{liu2023gender} were evaluated as being nicer and more pleasant to interact with while interacting with Amazon’s Alexa VoiceBot than their male counterparts. Similarly, female participants in \cite{kramer2010know} evaluated an embodied agent more favorably than male participants. These results suggest that users’ gender identity could influence people’s responses to technology and media \citep{brunel2003message, cai2017gender, darley1995gender, deaux1984individual}. 

Regarding the effect of similarity on gaze and snack selection, we found that opposite-gender pairings actually showed potential to enhance outcomes. For gaze ratio, male participants looked more at the female VR-ECA than the male VR-ECAs during the conversation and female participants looked more at the male VR-ECAs than their male counterparts. In addition, participants demonstrated slight trend of selecting a healthy snack after the conversation if the VR-ECA was the opposite gender (though the results were not statistically significant). While few studies have examined the influence of dissimilarity, \cite{zanbaka2006can} showed that interacting with virtual speakers of the opposite gender improved attitudes toward the topics of persuasive messages. Participants in \cite{carli1990gender} also reported that the opposite-gender (human) speakers had more influence. These results suggest that factors other than similarity influenced outcomes. While more research is needed to fully unpack and replicate these results, we offer a few potential explanations about why gender matching did not elicit expected outcomes. One explanation relates to social scripts, or social and biological factors related to attraction and mate selection. Although participants were certainly aware of the fact that they were interacting with an AI agent, the visual embodiment created a very immediate impression that may have steered automatic social behaviors. For instance, both male as well as female agents were generally considered attractive, having a symmetric face, clear skin, and a well-proportioned body in line with gender-typical physical attractiveness cues (e.g. man: strong jaw, muscular body; woman: full lips and high waist-to-hip ratio). Thus, given that gaze behavior is largely implicit and beyond conscious control, participants might have been more interested in, and attentive to the other gender. This effect would run counter to our similarity manipulation. Also, participants would become more health conscious to appear more attractive to the opposite gender. This biological response would run counter to the intended similarity manipulation. 

Another explanation could be that contextual factors beyond the direct human-to-agent interaction could have influenced the effect of gender matching. For example, the consultation task involved task-focused dialogue (discussing nutrition and exercise habits) rather than social dialogue \citep{bickmore2005social}. \cite{vugt2008effects} found people’s responses to facially similar embodied agents varied by the agents’ helpfulness. Thus, it is possible that people’s perceptions of the helpfulness of virtual health coaches obscured the effects of gender matching on outcomes. In addition, the participants engaged in back-and-forth conversations with the virtual AI agent for 5 minutes. This means that interpersonal interaction and linguistics-related factors, such as utterance sequences \citep{solomon2021dynamic} and turn-taking patterns \citep{levinson2016turn, sacks1978simplest, skantze2021turn}, could have moderated the effect of gender matching on outcomes. 

\subsection{Implications for Communication Research }
\subsubsection{Artificial Influence: A New Field of Research about How AI Can Change Human Behavior}
The embodied virtual AI paradigm expands our understanding of artificial influence, a rapidly expanding field at the intersection of interpersonal communication, persuasion, and AI. Prior research related to AI, virtual agents, and persuasion largely focused on one particular modality - text-only conversations. (e.g., the effect of text-based messages generated by large language models; \cite{lim2023artificial, karinshak2023working}). Alternatively, prior work only studied human-agent interaction without natural language generation capabilities \citep{liaw2023artificial}. The current work combines the verbal, nonverbal, and experience of physically being in the environment, which better mimics human-to-human interactions that shaped the foundation of social influence research. In addition, it opens the door to investigating persuasive communication in highly controlled experimental settings, circumventing the noise that is inherent in real-life interactions or other tasks currently used in interpersonal communication (e.g., using confederates, which is often impossible, implausible, and always expensive). 

While the strength of this paradigm may at first appear more as a methodological advancement, we argue that this advancement is necessary for deeper theoretical insights into the mechanisms of influence, particularly the mechanisms of similarity that were the focus of this investigation. For instance, existing literature overwhelmingly supports the notion that similarity is a central ingredient of persuasion and social influence, as illustrated by homophily and social identity theory. However, many studies of homophily use computational methods, social network analysis, and other macro-level approaches \citep{mcpherson2001birds, kossinets2009origins}. However, treating individuals with certain characteristics as nodes in similarity-based networks overlooks the detailed communicative processes that occur during human-to-human - or now in human-to-AI interactions. Furthermore, other studies examining similarity within persuasion tend to focus on perceived similarity from one-way communication (e.g., reading narratives whose protagonists are similar to the reader; \cite{andsager2006perceived}; listening to or watching recordings of messages; \cite{kim2016effect}) or outcomes from one-time interactions where first impressions matter \citep{ahn2021ai}. Influence, on the other hand, both human-to-human and AI-to-human, is a dynamic process that involves what is being said (verbal message), as well as how it’s being said (nonverbals), and other factors of the interacting parties (e.g., observable similarity, sense of rapport).

Our research paradigm, then, augments existing research by illustrating how the persuasive power of factors such as interpersonal similarity can vary based on the context (e.g., conversation about health vs. social dialogue; one-time vs. longitudinal interaction) and other variables. With this in mind, it is also clear that our analyses have barely scratched the surface of what is possible in terms of studying the actual interaction processes between humans and embodied AIs - in an over-time fashion and simultaneously monitoring effects on multiple social behaviors (e.g. gaze, facial expressions, verbal content, etc.).

\subsection{Strengths, Limitations, and Future Research}
As with all research, our study balances several strengths and limitations that are worth considering when interpreting the results. Regarding strengths, we can refer to the many positive comments received by participants, who generally expressed a willingness to interact with the VR-ECA, were surprised that this AI embodiment was already possible, and saw the immense potential for practical applications. 

However, although participants were overall quite enthusiastic about the study, it also surfaced some of the remaining technical challenges. For example, one key limitation refers to the turn-taking speed, which in our study was on the order of seconds (the time it takes for TTS, GPT-API querying, and STT re-conversion of the retrieved LLM results). Going forward, it will be important to reduce this lag to a time range that resembles the speed of natural human-human conversations (around 200 milliseconds /cite{stivers2009universals}). A second limitation relates to the quality of our embodiment, specifically the visual fidelity and fine detail of the agents. Although the agents resembled males and females in ways that participants were able to understand, their visual appearance was still somewhat comic-like. Given that leading players in the VR industry (e.g. Meta, Unreal, Apple) have already proposed high-realism avatars that almost perfectly look like humans, it is only a question of time until we can study the influence of these characteristics. That said, participants did perceive the used agents in the intended ways, and future work may even explore alternative appearances, such as figures with whom participants have parasocial relationships to leverage further influence. Next, our agents only exhibited rudimentary nonverbals (like speech-lip synchronization and some body-sway) that were less flexible and dynamic than the language generation by LLMs. Finally, in natural human-human communication, the nonverbal and the verbal behavior streams are often co-dependent (e.g. using gestures or facial expressions to add emphasis to text, or using nonverbals like nodding for backchanneling), but this was not realized in the current setup. Technically, such an integration is on the verge of what is possible, and we can certainly expect major progress in this area. 

\section{Conclusion}
To summarize, we examined the influence of gender-matched VR-embodied conversation agents (ECA). This was achieved by combining the potential of VR to simulate human-like personae (with a physical appearance, a voice, and lively nonverbal dynamics) with a state-of-the-art LLM-based AI agent. Regarding the impact of embodiment and VR, we find that participants experienced greater immersion and presence after interacting with VR-ECA vs. text-based CAs. The similarity manipulation showed that female participants liked the VR-ECA more than male participants, regardless of the agents’ gender. In addition, opposite-gender VR-ECAs showed the potential to enhance overt attention and healthy snack selection. Overall, we expect that the massive expansion of AI-LLMs is soon going to merge with VR, creating a fully embodied human-like simulacra. As this happens, current text-based large-language models (tb-LLMs) will evolve into multimodal communication models (MCMs) that feature the full spectrum of social embodiment. Although such a development will certainly have social consequences beyond our current research, from a research standpoint, this will enable us to examine a myriad of questions related to persuasion and social interaction - far beyond the basic gender-matching effects studied here. 

\section*{Data Availability Statement}
This study’s data are available on Github, at [Anonymized for review].

\bibliographystyle{elsarticle-harv}
\bibliography{main}

\end{document}